\documentclass[smallextended]{svjour3}       
\smartqed  
\usepackage{graphicx,epsfig}

\journalname{Quantum Inf. Process.}

\begin{document}

\title{
Behavior of quantum discord, local quantum uncertainty, and local quantum Fisher
information in two-spin-1/2 Heisenberg chain with DM and KSEA interactions
}

\author{Anna~V.~Fedorova \and M.~A.~Yurischev*
}


\institute{A.~V.~Fedorova \at 
Institute of Problems of Chemical Physics, Russian Academy of Sciences,
Chernogolovka 142432, Moscow Region, Russia\\     
              \email{panna@icp.ac.ru}           
           \and
M.~A.~Yurischev \at
Institute of Problems of Chemical Physics, Russian Academy of Sciences,
Chernogolovka 142432, Moscow Region, Russia\\     
\email{yur@itp.ac.ru}
}

\date{Received:}

\titlerunning{
Behavior of quantum discord, local quantum uncertainty, and ...
}
\maketitle

\begin{abstract}
A two-qubit Heisenberg XYZ model with Dzyaloshinsky--Moriya (DM) and
Kaplan--Shekhtman--Entin-Wohlman--Aharony (KSEA) interactions is considered at thermal
equilibrium.
Analytical formulas are derived for the local quantum uncertainty (LQU) and
local quantum Fisher information (LQFI).
Using the available expressions for the entropic quantum discord, we perform a
comparative study of these measures of nonclassical correlation.
Our analysis showed the following:
all three measures of quantum correlation have similar qualitative and even
quantitative behavior on temperature for different values of system parameters,
there are regions in the parameter space which correspondent to a local increase of
correlations with increasing temperature, and
sudden changes in the behavior of quantum correlations occur at certain values of the
interaction parameters.
\end{abstract}

\keywords{
\and Heisenberg spin model
\and Density matrix
\and Quantum correlations
\and Discord
\and Local quantum uncertainty
\and Quantum Fisher information
}

\section{Introduction}
\label{sect:Intro}
The concept of quantum information correlation is central to modern quantum
information science.
Until the 21st century, quantum correlation meant entanglement.
It manifests itself in the Einstein-Podolsky-Rosen gedanken (thought) experiment,
Bell's inequality test, quantum cryptography, superdense coding, teleportation, etc.
\cite{P98,NC00} (see also review articles \cite{GRTZ02,AFOV08,HHHH09,RDBCLBAL09}).
Quantum correlations are
considered as a
physical resource (``as real as
energy'' \cite{HHHH09})\footnote{
 However, ``it is important to realize that in physics today, we have no knowledge of
 what the energy {\em is}'' \cite {FLS64}.
}.
This statement requires a
mathematical definition, since,
according to Kant,
``in jeder besonderen Naturlehre nur so viel eigentliche Wissenschaft angetroffen
werden k$\rm\ddot o$nne, als darin Mathematik anzutreffen ist''
(``in any special doctrine of nature there can be only as much {\em proper\/} science
as there is {\em mathematics} therein'' or
``every natural science contains as much truth as much mathematics it contains'').
Quantum entanglement was quantified in 1996, first for pure states
\cite{BBPSSW96,BBPS96}, and then for mixed states \cite{BDSW96}.
According to the accepted definition, the entanglement of a bipartite pure state is
the von Neumann entropy either of the two subsystems.\footnote{
 Earlier, a similar definition was proposed by Everett for the
 {\em canonical correlation} \cite{E73}.
}
The entanglement (of formation) of a bipartite mixed state is defined as the
minimum entanglement of an ensemble over all ensembles realizing the mixed state.

At one time it was believed that quantum entanglement is the main ingredient of
quantum speedup in quantum computation and communication, but there was no strong
evidence.
Moreover, in 1998, Knill and Laflamme showed, using the model of deterministic quantum
computation with one pure qubit (DQC1) \cite{KL98}, that computation can achieve
an exponential improvement in efficiency over classical computers even without
containing much entanglement.

In 2000-2001, {$\rm\dot Z$}urek et al. developed the concept of quantum discord -- ``a
measure of the quantumness of correlations'' \cite{Z00,OZ01}.
Simultaneously and independently, Vedral et al. \cite{HV01,V03} proposed a measure for
the purely classical correlation, which, after subtracting it from the total
correlation, led to the same amount of quantum correlation as the discord.
Then Datta et al. \cite{D08,DSC08} calculated discord in the Knill-Laflamme DQC1 model
and showed that it scales with the quantum efficiency, while entanglement remains
vanishingly small throughout the computation.
This attracted a lot of attention to the new measure of quantum correlation
\cite{M11,MBCPV11,MBCPV12,AFY14,S15}.

Quantum discord and entanglement are the same for the pure quantum states.
However discord can exist in separable mixed states, i.e., when quantum entanglement
is identically equal to zero.
The set of separable states possesses a nonzero volume in the whole Hilbert space of a
system \cite{ZHSL98} (it is a necessary condition for the arising of entanglement
sudden death (ESD) effect \cite{YE09}), whereas the set of states with zero discord,
vise versa, is negligibly small \cite{FACCA10}.
This circumstance alone sharply distinguishes discord from entanglement.
Moreover, numerous theoretical and experimental investigations of different
quantum system have clearly shown that while the quantum entanglement and discord
measure the same think -- the quantum correlation, but as a matter a fact,
discrepancies in quantitative and even qualitative behavior are very large
\cite{WR10,GMZS11,CRLB13}.
Discord and entanglement behave differently even for simplest mixed states --- the
Werner and Bell-diagonal ones (see, e.g, \cite{MGY17}).
This has led many to talk about entanglement and discord as different types of quantum
correlations.

However, the subsequent proposals with more and more new measures of quantum
correlations \cite{ABC16,BDSRSS18} caused a dilemma: should each measure be attributed
to its own correlation, or should it be argued that there is only one quantum
correlation, but
the methods of describing it are not adequate?
The physicists community now prefers to talk about entanglement and discord-like
quantum correlations \cite{BT17}.
As is customary for brevity, we will also refer to the various measures of quantum
correlation as ``quantum correlations''.
Nevertheless, the quantum correlation is one, but now there are only different
measures of it, which are still imperfect.
We would like to have such measures of quantum correlations that would be useful for
estimating the values of the speedup of quantum computing, the efficiency of quantum
heat engines, etc.
Let the measures be different, but the results must be the same.
(As, for example, there are various formulations of quantum mechanics which differ
dramatically in mathematical and conceptual overview, yet each one makes identical
predictions for all experimental results \cite{SBB02}.)
%

In the present paper we study the behavior simultaneously of three measures of quantum
correlation: entropic quantum discord, local quantum uncertainty, and local quantum
Fisher information (definitions for them are given in the next section).
Calculations are carried out using a fully anisotropic Heisenberg model of two
spin-1/2 with taken into account the Dzyaloshinsky--Moriya (DM) and
Kaplan--Shekhtman--Entin-Wohlman--Aharony (KSEA) interactions.
The model is considered in thermal equilibrium with a thermal bath.
Through extensive graphical analysis, we find that the behavior of these
significantly different measures demonstrate a similar qualitative and, in many cases,
acceptable quantitative agreement with each other.

The organization of this paper is as follows.
We begin in Sect.~\ref{sect:prelim} with a brief overview of the quantum correlation
measures used in our work.
The model is described in Sect.~\ref{sect:H-rho}.
Expressions for the quantum correlations are derived and presented in
Sect.~\ref{sect:Express}.
Section~\ref{sect:Discuss} is devoted to a detail description and discussion of
different effects in behavior of quantum correlations under question.
Our main conclusions are summarized in Sect.~\ref{sect:Concl}.

\section{
Preliminaries
}
\label{sect:prelim}
Here we recall some notions and equations that will be needed in the following
sections.
 
\subsection{
Quantum discord
}
\label{sect:QD}
The entropic quantum discord $Q$ for a bipartite quantum state $\rho$ is defined as
the minimum difference between two classically-equivalent expressions of the mutual
information \cite{OZ01}: $Q(\rho)=I-J$, where $I$ is the usual quantum mutual
information and $J$ the local measurement-induced quantum mutual information.
Below we will deal with the Bell-diagonal states.
Exact explicit formula for the quantum discord of these quantum states has been
derived by Luo \cite{Luo08}.
Notice that another quantity of quantum correlation, namely the one-way quantum work
deficit coincides the quantum discord in Bell-diagonal states (see, e.g.,
\cite{YF16}).

\subsection{
Local quantum uncertainty
}
\label{sect:LQU}
The local quantum uncertainty (LQU) as a measure of quantum correlation, $\cal U$, was
appeared in 2013~\cite{GTA13}  (see also \cite{KHPD21} and references therein).
It is defined as the minimum quantum uncertainty associated to a single measurement on
one subsystem, say $A$, of bipartite system $AB$.
The authors \cite{GTA13} have evaluated this measure in the case of $2\times d$
systems:
\begin{equation}
   \label{eq:U}
   {\cal U}(\rho)=1-\lambda_{max}(W),
\end{equation}
where $\lambda_{max}$ denotes the maximum eigenvalue of the $3\times3$ symmetric
matrix $W$ whose entries are
\begin{equation}
   \label{eq:W}
   W_{\mu \nu}={\rm Tr}\{\rho^{1/2}(\sigma_\mu\otimes{\rm I})\rho^{1/2}(\sigma_\nu\otimes{\rm I})\}
\end{equation}
with $\mu,\nu=x,y,z$ and $\sigma_{x,y,z}$ are the Pauli matrices.

\subsection{
Local quantum Fisher information
}
\label{sect:LQFI}
Fisher's concept of information \cite{Fisher1925} has a long history and wide
applications \cite{E98,PG10,LMVGW17,MCWV11,JA20}.
A measure  of nonclassical correlations based on it was suggested in
2014~\cite{GSGTFSSOA14} (see there especially Supplementary Information) under
name ``interferometric power''; see also \cite{KLKW18}.
This measure which we will denote by $\cal F$ equals the optimal local quantum Fisher
information (LQFI) $F$ with the measuring operator $H_A$ acting in the subspace of
party $A$:
\begin{equation}
   \label{eq:F}
   {\cal F}(\rho)=\min_{H_A}F(\rho,H_A).
\end{equation}

\section{
Hamiltonian and density matrix
}
\label{sect:H-rho}
Consider a two-qubit fully anisotropic Heisenberg model with DM and KSEA interactions
\cite{Y20}.
In a zero external field, Hamiltonian reads
\begin{eqnarray}
   \label{eq:H}
   {\cal H}=J_x\sigma_1^x\sigma_2^x + J_y\sigma_1^y\sigma_2^y + J_z\sigma_1^z\sigma_2^z
	 +D_z(\sigma_1^x\sigma_2^y-\sigma_1^y\sigma_2^x)
	 +{\rm \Gamma}_z(\sigma_1^x\sigma_2^y+\sigma_1^y\sigma_2^x).\ 
\end{eqnarray}
Its matrix form has the X structure:
\begin{equation}
   \label{eq:HH}
   {\cal H}=
	 \left(
      \begin{array}{cccc}
      J_z&.&.&J_x-J_y-2i{\rm \Gamma}_z\\
      .&-J_z\ &J_x+J_y+2iD_z&.\\
      .&J_x+J_y-2iD_z\ &-J_z&.\\
      J_x-J_y+2i{\rm \Gamma}_z&.&.&J_z
      \end{array}
   \right),
\end{equation}
where the dots are put instead of zero entries.
The energy levels are given as
\begin{equation}
   \label{eq:Ei}
   E_{1,2}=J_z\pm r_1,\qquad E_{3,4}=-J_z\pm r_2,
\end{equation}
where
\begin{equation}
   \label{eq:r1r2}
   r_1=[(J_x-J_y)^2+4{\rm\Gamma}_z^2]^{1/2},\qquad
   r_2=[(J_x+J_y)^2+4D_z^2]^{1/2}.
\end{equation}
Note that ${\rm\Gamma}_z$ (constant of KSEA interaction) is accumulated only in $r_1$,
while the constant of DM coupling, $D_z$, is contained entirely in the coefficient
$r_2$.

The partition function $Z=\sum_i\exp(-\beta E_i)$ equals
\begin{equation}
   \label{eq:ZT}
   Z=2(e^{-\beta J_z}\cosh\beta r_1+e^{\beta J_z}\cosh\beta r_2),
\end{equation}
where
$\beta=1/T$ and $T$ is the absolute temperature in energy units.
The Gibbs density matrix is given as
\begin{equation}
   \label{eq:rhoG}
   \rho=\frac{1}{Z}\exp(-\beta{\cal H}).
\end{equation}
Calculations yield
\begin{eqnarray}
   \label{eq:rho}
   \rho=
	 \left(
      \begin{array}{cccc}
      a&.&.&u\\
      .&b\ &v&.\\
      .&v^*\ &b&.\\
      u^*&.&.&a
      \end{array}
   \right)
\end{eqnarray}
(the asterisk denotes complex conjugation).
Here
\begin{eqnarray}
   \label{eq:avT}
   &&a=\frac{1}{Z}e^{-\beta J_z}\cosh\beta r_1,\quad
   u=-\frac{1}{Z}\frac{J_x-J_y-2i{\rm\Gamma}_z}{r_1}e^{-\beta J_z}\sinh\beta r_1,
   \nonumber\\
   &&b=\frac{1}{Z}e^{\beta J_z}\cosh\beta r_2,\quad
   v=-\frac{1}{Z}\frac{J_x+J_y+2iD_z}{r_2}e^{\beta J_z}\sinh\beta r_2,
\end{eqnarray}
where $r_1$ and $r_2$ are given again by Eq.~(\ref{eq:r1r2}).

Using the invariance of quantum correlations under any local unitary transformations
(see, for example, \cite{MBCPV12}), we remove complex phases in the off-diagonal
entries and change $\rho\to\varrho$, where
\begin{eqnarray}
   \label{eq:varrho}
   \varrho=
	 \left(
      \begin{array}{cccc}
      a&.&.&|u|\\
      .&b\ &|v|&.\\
      .&|v|\ &b&.\\
      |u|&.&.&a
      \end{array}
   \right)
\end{eqnarray}
with
\begin{equation}
   \label{eq:uv}
   |u|=\frac{1}{Z}e^{-\beta J_z}\sinh\beta r_1,\qquad
   |v|=\frac{1}{Z}e^{\beta J_z}\sinh\beta r_2.
\end{equation}
Via orthogonal transformation
\begin{equation}
   \label{eq:R}
   R=\frac{1}{\sqrt{2}}
	 \left(
      \begin{array}{ccrr}
      1&.&.&1\\
      .&1&1&.\\
      .&1&-1&.\\
      1&.&.&-1
      \end{array}
   \right)=R^t
\end{equation}
(the subscript $t$ stands for matrix transpose),
the density matrix $\varrho$ is reduced to the diagonal form 
\begin{eqnarray}
   \label{eq:RvarrhoR}
   R\varrho R=
	 \left(
      \begin{array}{cccc}
      p_1&.&.&.\\
      .&p_2\ &.&.\\
      .&.\ &p_3&.\\
      .&.&.&p_4
      \end{array}
   \right),
\end{eqnarray}
where eigenvalues equal
\begin{equation}
   \label{eq:p_i}
   p_1=a+|u|,\quad
	 p_2=b+|v|,\quad
	 p_3=b-|v|,\quad
	 p_4=a-|u|.
\end{equation}
The corresponding eigenvectors of $\varrho$ are given as
\begin{equation}
   \label{eq:v1-4}
   |1\rangle=
   \frac{1}{\sqrt2}
	 \left(
      \begin{array}{c}
      1\\
      .\\
      .\\
      1
      \end{array}
   \right),\
   |2\rangle=
   \frac{1}{\sqrt2}
	 \left(
      \begin{array}{c}
      .\\
      1\\
      1\\
      .
      \end{array}
   \right),\ 
   |3\rangle=
   \frac{1}{\sqrt2}
	 \left(
      \begin{array}{c}
      .\\
      1\\
      -1\\
      .
      \end{array}
   \right),\ 
   |4\rangle=
   \frac{1}{\sqrt2}
	 \left(
      \begin{array}{c}
      1\\
      .\\
      .\\
      -1
      \end{array}
   \right).
\end{equation}
These are the Bell vectors $|\Phi^+\rangle$, $|\Psi^+\rangle$,
$|\Psi^-\rangle$, and $|\Phi^-\rangle$, respectively.

\section{
Expressions for the quantum correlations
}
\label{sect:Express}
The state (\ref{eq:varrho}) belongs to the Bell-diagonal family which in turn is a
subclass of X quantum states.
It is noteworthy that both entropic quantum discord and one-way quantum work deficit
give the same results not only for the Bell diagonal states, but even for the X
quantum states if the marginal state of one qubit is maximally mixed and measurements
are performed on this qubit \cite{YF16}.

\subsection{
Quantum discord
}
\label{subsect:Q}
Quantum discord in the case of Bell diagonal states can be written as
\cite{Y15,FWBAC10}
\begin{equation}
   \label{eq:Q0Q1}
   Q=\min\{Q_0,Q_1\}.
\end{equation}
The branch $Q_0$ corresponds to the zero optimal measurement angle and is given as
\begin{equation}
   \label{eq:Q0}
   Q_0=-S-2(a\log_2a+b\log_2b),
\end{equation}
where $a$ and $b$ are determined by Eq.~(\ref{eq:avT}) and $S$ is the entropy of the
system in bits:
\begin{eqnarray}
   \label{eq:S}
   &&S\equiv-\sum_{i=1}^4p_i\log_2p_i=\log_2Z
   \nonumber\\
	 &&-\frac{2\beta}{Z\ln2}[e^{-\beta J_z}(r_1\sinh\beta r_1-J_z\cosh\beta r_1)
	 +e^{\beta J_z}(r_2\sinh\beta r_2+J_z\cosh\beta r_2)].\quad\ \ 
\end{eqnarray}
The second branch $Q_1$ corresponds to the $\pi/2$ optimal measurement angle and is
expressed as
\begin{equation}
   \label{eq:Q1}
   Q_1=1-S-\frac{1+w}{2}\log_2\frac{1+w}{2}-\frac{1-w}{2}\log_2\frac{1-w}{2},
\end{equation}
where
\begin{equation}
   \label{eq:w}
   w=2(|u|+|v|)=\frac{2}{Z}(e^{-\beta J_z}\sinh\beta r_1+e^{\beta J_z}\sinh\beta r_2).
\end{equation}
The transition threshold from one branch to another is determined by the equation
$Q_0=Q_1$ or in open form,
\begin{equation}
   \label{eq:Q-bond}
   \ln2+2(a\ln a+b\ln b)-\frac{1+w}{2}\ln\frac{1+w}{2}-\frac{1-w}{2}\ln\frac{1-w}{2}=0.
\end{equation}

\subsection{
Optimal LQU
}
\label{subsect:LQU}
Using transformation (\ref{eq:R}) we get matrix elements
$\langle m|\sigma_\mu\otimes I|n\rangle$ in the diagonal representation of the density
matrix $\varrho$:
\begin{eqnarray}
   \label{eq:RxR}
   R(\sigma_x\otimes I)R=
	 \left(
      \begin{array}{cccc}
      .&1&.&.\\
      1&.\ &.&.\\
      .&.\ &.&-1\\
      .&.&-1&.
      \end{array}
   \right),
\end{eqnarray}
\begin{eqnarray}
   \label{eq:RyR}
   R(\sigma_y\otimes I)R=
	 \left(
      \begin{array}{cccc}
      .&.&i&.\\
      .&.&.&i\\
      -i&.&.&.\\
      .&-i&.&.
      \end{array}
   \right),
\end{eqnarray}
\begin{eqnarray}
   \label{eq:RzR}
   R(\sigma_z\otimes I)R=
	 \left(
      \begin{array}{cccc}
      .&.&.&1\\
      .&.&1&.\\
      .&1&.&.\\
      1&.&.&.
      \end{array}
   \right).
\end{eqnarray}
From here, it is easy to see that the matrix $W$ defined by Eq.~(\ref{eq:W}) is
diagonal and its eigenvalues are equal to (for a comparison, see, e.g.,
\cite{JBD17,GPTTC21})
\begin{eqnarray}
   \label{eq:Wxx}
   W_{xx}=2(\sqrt{p_1p_2}+\sqrt{p_3p_4})
	 =2(\sqrt{(a+|u|)(b+|v|)}+\sqrt{(a-|u|)(b-|v|)}),\quad
\end{eqnarray}
\begin{eqnarray}
   \label{eq:Wyy}
   W_{yy}=2(\sqrt{p_1p_3}+\sqrt{p_2p_4})
	 =2(\sqrt{(a+|u|)(b-|v|)}+\sqrt{(a-|u|)(b+|v|)}),\quad
\end{eqnarray}
\begin{eqnarray}
   \label{eq:Wzz}
   W_{zz}=2(\sqrt{p_1p_4}+\sqrt{p_2p_3})
	 =2(\sqrt{a^2-|u|^2}+\sqrt{b^2-|v|^2}).
\end{eqnarray}
In explicit form
\begin{equation}
   \label{eq:Wxyz}
   W_{xx}=\frac{4}{Z}\cosh[\beta(r_1+r_2)/2],\quad
   W_{yy}=\frac{4}{Z}\cosh[\beta(r_1-r_2)/2],\quad
   W_{zz}=\frac{4}{Z}\cosh\beta J_z.
\end{equation}
It is clear that $W_{xx}\ge W_{yy}$.
Therefore, the value of quantum correlation trough LQU equals
\begin{eqnarray}
   \label{eq:min_U}
   {\cal U}=\min\{{\cal U}_0,{\cal U}_1\},
\end{eqnarray}
where
\begin{eqnarray}
   \label{eq:U0U1}
   {\cal U}_0=1-W_{zz},\qquad
   {\cal U}_1=1-W_{xx}.
\end{eqnarray}
 
\subsection{
Optimal LQFI
}
\label{subsect:LQFI}
Local quantum Fisher information reads
\cite{GSGTFSSOA14,B14,SBDL19},\cite{H20}\footnote
   {Note that the expressions for the density matrix elements and partition function,
	 Eqs.~(23) and (24) in Ref.~\cite{H20},
   contain errors.
	 },
\cite{MC20,YLLF20}
\begin{equation}
   \label{eq:lqfi}
   F(\varrho,H_A)=\frac{1}{2}\sum_{m,n}\frac{(p_m-p_n)^2}{p_m+p_n}|\langle m|H_A|n\rangle|^2,
\end{equation}
where the operator $H_A$ acts in the subspace
of party $A$.
For qubit systems, one takes
\begin{equation}
   \label{eq:Ha}
   H_A=\vec\sigma\cdot\vec r
\end{equation}
with $|\vec r|=1$; $\vec\sigma=(\sigma_x,\sigma_y,\sigma_z)$ is the vector of the
Pauli matrices.
The relation (details can be found in \cite{B14,YLLF20})
\begin{equation}
   \label{eq:1F}
   \sum_{m\ne n}\frac{2p_mp_n}{p_m+p_n}|\langle m|H_A|n\rangle|^2
	 =\sum_{\mu,\nu=x,y,z}\sum_{m\ne n}\frac{2p_mp_n}{p_m+p_n}\langle m|\sigma_\mu\otimes I|n\rangle\langle n|\sigma_\nu\otimes I|m\rangle
\end{equation}
leads to ${\cal F}=1-\lambda_{max}$,
where $\lambda_{max}$ is the largest eigenvalue of the real symmetric $3\times3$
matrix $M$ with entries
\begin{equation}
   \label{eq:M}
   M_{\mu\nu}=\sum_{m\ne n}\frac{2p_mp_n}{p_m+p_n}\langle m|\sigma_\mu\otimes I|n\rangle\langle n|\sigma_\nu\otimes I|m\rangle.
\end{equation}
Using Eqs.~(\ref{eq:RxR})--(\ref{eq:RzR}), one finds that the matrix $M$ is also
diagonal and its nondiagonal elements (eigenvalues) are equal to
\begin{eqnarray}
   \label{eq:Mxx}
   M_{xx}=\frac{4p_1p_2}{p_1+p_2}+\frac{4p_3p_4}{p_3+p_4}
	 =\frac{4(a+|u|)(b+|v|)}{a+b+|u|+|v|}+\frac{4(a-|u|)(b-|v|)}{a+b-|u|-|v|},
\end{eqnarray}
\begin{eqnarray}
   \label{eq:Myy}
   M_{yy}=\frac{4p_1p_3}{p_1+p_3}+\frac{4p_2p_4}{p_2+p_4}
	 =\frac{4(a+|u|)(b-|v|)}{a+b+|u|-|v|}+\frac{4(a-|u|)(b+|v|)}{a+b-|u|+|v|},
\end{eqnarray}
\begin{eqnarray}
   \label{eq:Mzz}
   M_{zz}=\frac{4p_1p_4}{p_1+p_4}+\frac{4p_2p_3}{p_2+p_3}
	 =\frac{2(a^2-|u|^2)}{a}+\frac{2(b^2-|v|^2)}{b}.
\end{eqnarray}
In explicit form,
\begin{eqnarray}
   \label{eq:Mxyz}
   &&M_{xx}=\frac{4}{Z}\frac{e^{\beta J_z}\cosh\beta r_1+e^{-\beta J_z}\cosh\beta r_2}{\cosh 2\beta J_z+\cosh[\beta(r_1-r_2)]},
   \nonumber\\
   &&M_{yy}=\frac{4}{Z}\frac{e^{\beta J_z}\cosh\beta r_1+e^{-\beta J_z}\cosh\beta r_2}{\cosh 2\beta J_z+\cosh[\beta(r_1+r_2)]},\\
   &&M_{zz}=\frac{2}{Z}\frac{e^{\beta J_z}\cosh\beta r_1+e^{-\beta J_z}\cosh\beta r_2}{\cosh\beta r_1\cosh\beta r_2}.
   \nonumber
\end{eqnarray}
It is seen that $M_{xx}\ge M_{yy}$.
Hence, the value of the quantum correlation in terms of LQFI is defined by equation
\begin{equation}
   \label{eq:min_F}
   {\cal F}=\min\{{\cal F}_0,{\cal F}_1\},
\end{equation}
where
\begin{equation}
   \label{eq:F0F1}
   {\cal F}_0=1-M_{zz},\qquad
   {\cal F}_1=1-M_{xx}.
\end{equation}
 
\subsection{
Boundaries between branches
}
\label{subsect:boundary}
Equation for the boundary separating the regions with branches ${\cal U}_0$
and ${\cal U}_1$ is ${\cal U}_0={\cal U}_1$.
Using Eqs.~(\ref{eq:Wxyz}) and (\ref{eq:U0U1}) we get the solution
\begin{equation}
   \label{eq:bound}
   r_1+r_2=2|J_z|.
\end{equation}
In turn, equation for the boundary between two branches ${\cal F}_0$ and ${\cal F}_1$
is also reduced to the condition (\ref{eq:bound}).
Moreover, performing direct calculations (by hand or using the package Mathematica) it
is easy to prove that the transcendental equation (\ref{eq:Q-bond}) has a solution
$|J_z|=(r_1+r_2)/2$.
It is remarkable that the branches of the three measures under study are separated by
the same boundary (\ref{eq:bound}).

The formulas presented in this section open a way to investigate the behavior of
nonclassical correlations in the thermolyzed system (\ref{eq:H}).

%
\begin{figure}[t]
\begin{center}
\epsfig{file=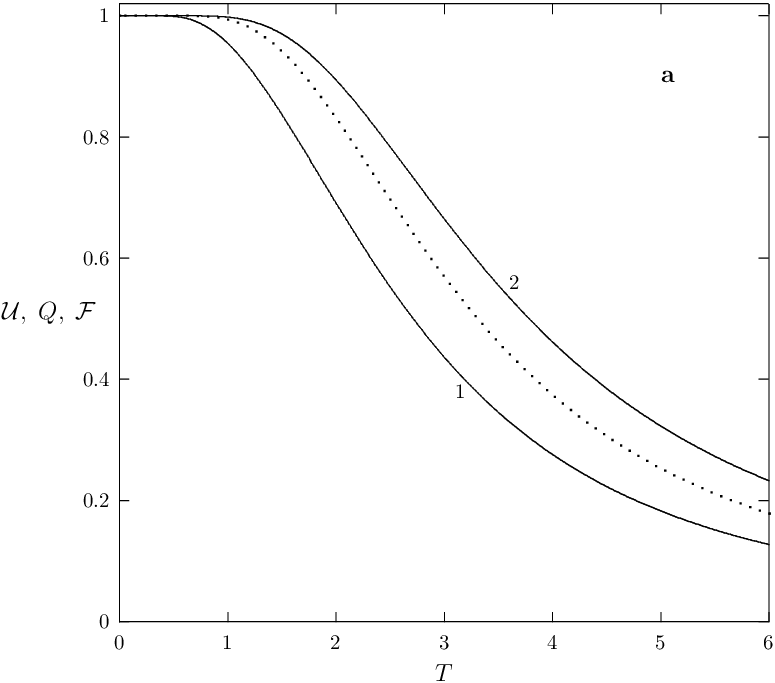,width=5.7cm}
\hspace{2mm}
\epsfig{file=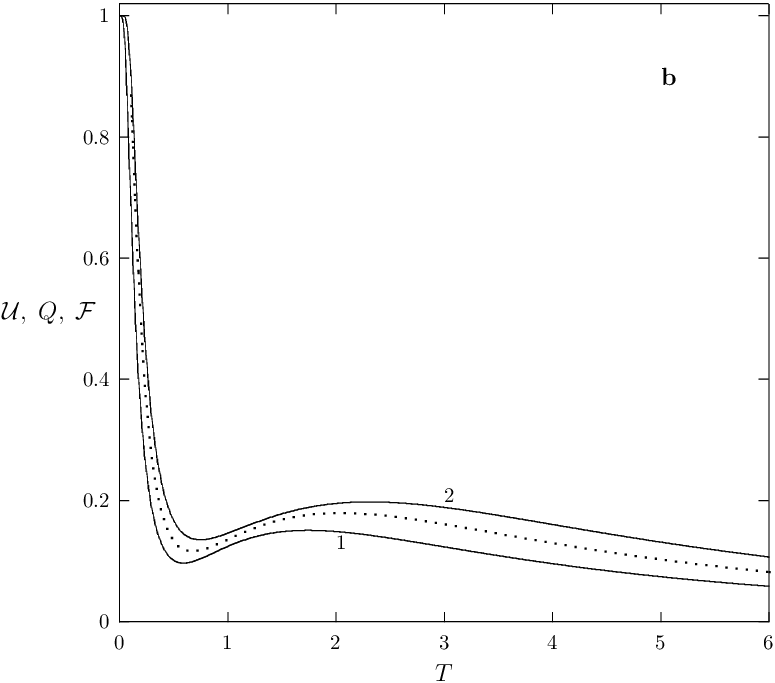,width=5.7cm}
\end{center}
\begin{center}
\caption{
Quantum correlations $\cal U$ (solid line 1), $Q$ (dotted line), and $\cal F$ (solid
line 2) versus temperature $T$ for $J_x=-1$, $J_y=-1.5$, $D=1.8$, ${\rm\Gamma}_z=0.3$
and $J_z=2$~($\bf a$) and $J_z=-2$~($\bf b$)
}
\label{fig:zUQFa}
\end{center}
\end{figure}
%
\section{
Results and discussion
}
\label{sect:Discuss}
Before starting a general analysis, consider two examples with different behavior of
quantum correlations.
Figure~\ref{fig:zUQFa}, (a) and (b), shows the dependencies of LQU, discord $Q$, and
LQFI as functions of temperature.
Interaction constants $J_x$, $J_y$, $D_z$, and ${\rm\Gamma}_z$ are the same in both
cases (a) and (b), while $J_z$ differs only in sign:
$J_z=2$ (antiferromagnetic exchange coupling) and $J_z=-2$ (ferromagnetic exchange
coupling).

One can see the following.
All curves go to zero as the temperature rises.
On the other hand, in the limit $T\to0$, quantum correlations reach the maximum
possible value equaling one and their first derivatives with respect to the
temperature equals zero at $T=0$.
This leads to quasi-horizontal sections on the curves.
Characteristic length (temperature) of these sections, $T_{ch}$, one could try to
relate with the energy gap $\Delta E$ in the spectrum: $T_{ch}\sim\Delta E$.
Using Eqs.~(\ref{eq:Ei}) and (\ref{eq:r1r2}) and numerical values of interaction
constants given in the figure caption we obtain the estimations: $T_{ch,a}\sim7.6$ for
the dependencies in Fig.~\ref{fig:zUQFa}a and $T_{ch,b}\sim0.4$ for the
dependencies in Fig.~\ref{fig:zUQFa}b.
Looking at the curves in the figure, we conclude that the estimates correctly give
that the quasi-horizontal section in the antiferromagnetic case is much larger
than in the ferromagnetic case.
However, as can be seen from Fig.~\ref{fig:zUQFa}, both estimates give an order of
magnitude overestimated values.

Next, the behavior of quantum correlations in Fig.~\ref{fig:zUQFa}a
characterized by {\em monotonic} decrease from one to zero.
We will refer to this behavior as type I behavior.
It is noteworthy that the curves shown in Fig.~\ref{fig:zUQFa}b, also decrease from
one to zero, but have local rise in the intermediate temperature range
(approximately from $T_1\approx0.6$ to $T_2\approx2.2$).
Such behavior with local minimum and maximum at $T>0$ will be referred to as type II.

It is seen from Fig.~\ref{fig:zUQFa} that the dependencies for temperatures
$T\in(0,\infty)$ satisfy the inequalities ${\cal U}<Q<{\cal F}$.
Finally, the curves for both $J_z=2$ and $J_z=-2$ repeat the behavior of each other
quite well.

\subsection{
High-temperature behavior
}
\label{subsect:Too}
The observed behavior at high temperatures can be confirmed by rigorous calculations.
Using formulas for the different quantum correlations derived in the previous section,
we obtain for the quantum discord
\begin{equation}
   \label{eq:Q0hT}
	 Q_0(T)|_{T\to\infty}=\frac{r_1^2+r_2^2}{4T^2\ln2}+\frac{J_z(r_2^2-r_1^2)}{4T^3\ln2}+O(1/T^4),
\end{equation}
\begin{equation}
   \label{eq:Q1hT}
	 Q_1(T)|_{T\to\infty}=\frac{4J_z^2+(r_1-r_2)^2}{8T^2\ln2}+\frac{J_z(r_2^2-r_1^2)}{4T^3\ln2}+O(1/T^4),
\end{equation}
for the LQU
\begin{equation}
   \label{eq:1WzzhT}
	 {\cal U}_0(T)|_{T\to\infty}=\frac{r_1^2+r_2^2}{4T^2}+\frac{J_z(r_2^2-r_1^2)}{4T^3}+O(1/T^4),
\end{equation}
\begin{equation}
   \label{eq:1WxxhT}
	 {\cal U}_1(T)|_{T\to\infty}=\frac{4J_z^2+(r_1-r_2)^2}{8T^2}+\frac{J_z(r_2^2-r_1^2)}{4T^3}+O(1/T^4),
\end{equation}
and for the LQFI
\begin{equation}
   \label{eq:1MzzhT}
	 {\cal F}_0(T)|_{T\to\infty}=\frac{r_1^2+r_2^2}{2T^2}+\frac{J_z(r_2^2-r_1^2)}{2T^3}+O(1/T^4).
\end{equation}
\begin{equation}
   \label{eq:1MxxhT}
	 {\cal F}_1(T)|_{T\to\infty}=\frac{4J_z^2+(r_1-r_2)^2}{4T^2}+\frac{J_z(r_2^2-r_1^2)}{2T^3}+O(1/T^4),
\end{equation}
Thus, quantum correlations decay at high temperatures according to the law $1/T^2$.

\subsection{
Local unitary transformation of $\varrho$ 
}
\label{subsect:O}
As mentioned above, quantum correlations are invariant under any local unitary
transformations.
Let us take a local unitary (orthogonal) transformation $O=I\otimes\sigma_x$,
\begin{equation}
   \label{eq:O}
   O=
	 \left(
      \begin{array}{rr}
      1&.\\
      .&1
      \end{array}
   \right)\otimes
	 \left(
      \begin{array}{rr}
      .&1\\
      1&.
      \end{array}
   \right)=
	 \left(
      \begin{array}{ccrr}
      .&1&.&.\\
      1&.&.&.\\
      .&.&.&1\\
      .&.&1&.
      \end{array}
   \right)=O^t.
\end{equation}
Using it, the density matrix (\ref{eq:varrho}) is transformed as follows:
\begin{eqnarray}
   \label{eq:OvarrhoO}
   O\varrho O=
	 \left(
      \begin{array}{cccc}
      b&.&.&|v|\\
      .&a\ &|u|&.\\
      .&|u|\ &a&.\\
      |v|&.&.&b
      \end{array}
   \right).
\end{eqnarray}
This means that quantum correlations do not change upon exchange
\begin{equation}
   \label{eq:Jz_r1r2}
   \{J_z,r_1,r_2\}\leftrightarrow\{-J_z,r_2,r_1\}.
\end{equation}
Thus, to describe all situations, it suffices to consider only the cases $J_z=0$ and
$J_z>0$ for different values $r_1$ and $r_2$
(results for $J_z<0$ will follow from results for $J_z>0$ with simultaneous
replace $r_1\rightleftharpoons r_2$).
We will consider both of these cases separately.
 
\subsection{
Case $J_z=0$
}
\label{subsect:Jz0}
Let us start with the zero value of the $J_z$ interaction constant.
Here the ground state energy is $E_0=-\min\{r_1, r_2\}$ and the energy gap equals
$|r_1-r_2|$.
Formulas for quantum correlations are greatly simplified.
For instance,
\begin{equation}
   \label{eq:Jz0U}
   {\cal U}=1-{\rm sech}[(r_1-r_2)/2T]
\end{equation}
and
\begin{equation}
   \label{eq:Jz0F}
   {\cal F}=\tanh^2[(r_1-r_2)/2T].
\end{equation}
The values of quantum correlations depend only on the relative distance $|r_1-r_2|$ in
the range for $r_1$ and $r_2$ from zero to infinity.

\begin{figure}[t]
\begin{center}
\epsfig{file=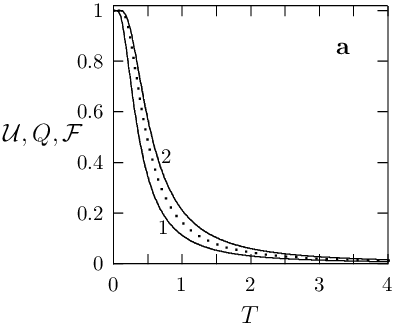,width=3.5cm}
\hspace{1mm}
\epsfig{file=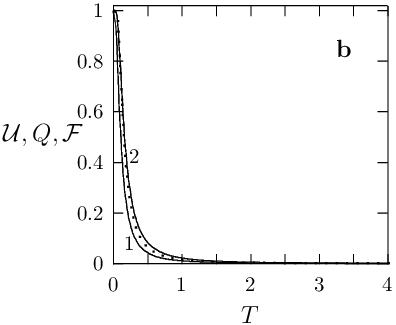,width=3.5cm}
\hspace{1mm}
\epsfig{file=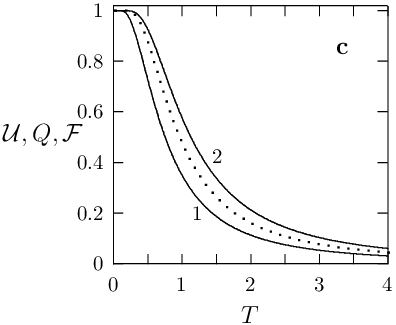,width=3.5cm}
\end{center}
\begin{center}
\caption{
Quantum correlations $\cal U$ (solid line 1), $Q$ (dotted line), and $\cal F$  (solid
line 2) as functions of temperature $T$ for $J_z=0$, $r_1=1$, and $r_2=0~(\bf a)$,
$0.7~(\bf b)$, and $3~(\bf c)$
}
\label{fig:z010-3}
\end{center}
\end{figure}

In fact, this case contains only one independent parameter.
Without loss of generality, we put $r_1=1$ ($r_1=1$ will play the role of a
normalization constant).
The dependencies of quantum correlations are drawn in Fig.~\ref{fig:z010-3}.
%
%
It can be seen from this figure that the curves have a monotonically decreasing shape
(refer to type I in our classification).
When the single parameter $r_2$ increases from zero to one [see
Fig.~\ref{fig:z010-3}~(a) and (b)], the values of the quantum correlations decrease at
given temperatures and completely vanish at $r_2=1$.

Indeed, for $r_2=1$ the energy spectrum, Eq.~(\ref{eq:Ei}), consists of two levels
$E_{1,2}=\pm1$, which are both two-fold degenerate.
At this point, the density matrix (\ref{eq:varrho}) takes the form
\begin{eqnarray}
   \label{eq:varrho0}
   \varrho_0=
	 \left(
      \begin{array}{cccc}
      a&.&.&|u|\\
      .&a\ &|u|&.\\
      .&|u|\ &a&.\\
      |u|&.&.&a
      \end{array}
   \right).
\end{eqnarray}
After the local unitary (orthogonal) transformation $H_2=H\otimes H$, where
\begin{eqnarray}
   \label{eq:Had}
   H=\frac{1}{\sqrt{2}}
	 \left(
      \begin{array}{cr}
      1&1\\
      1&-1
      \end{array}
   \right)
\end{eqnarray}
is the Hadamard transform, the density matrix (\ref{eq:varrho0}) is reduced to
diagonal form: $H_2\varrho_0 H_2={\rm diag}(a+|u|, a-|u|, a-|u|, a+|u|)$.
This means that the state (\ref{eq:varrho0}) is classical and therefore all
quantum correlations disappear.
The latter is also seen from Eqs.~(\ref{eq:Jz0U}) and (\ref{eq:Jz0F}).

With a further increase in $r_2$, quantum correlations revive again, as seen in
Fig.~\ref{fig:z010-3}c.

\subsection{
Case $J_z\ne0$
}
\label{subsect:Jneq0}
Taking $J_z$ as a normalized constant and setting it equal to unity, the problem for
the dependencies of quantum correlations on the dimensionless temperature $T$
will contain two independent parameters $r_1$ and $r_2$.
The functions $Q$, $\cal U$, and $\cal F$ are piecewise because each of them consist
of two branches.

\begin{figure}[t]
\begin{center}
\epsfig{file=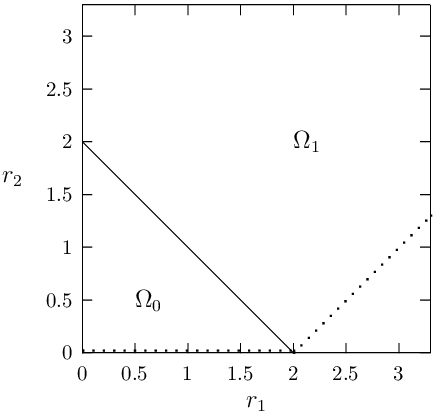,width=5.5cm}
\hspace{4mm}
\epsfig{file=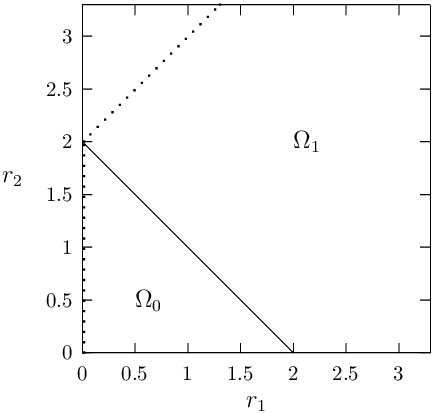,width=5.5cm}
\end{center}
\begin{center}
\caption{
Phase diagrams in the parameter space $(r_1,r_2)$ for $J_z=1$ (left) and $J_z=-1$
(right).
Solid line $r_1+r_2=2$ separates the regions $\Omega_0$ and $\Omega_1$.
Dotted broken lines
[except points
$(2,0)$ and $(0,2)$ in
the left and right panels, respectively] show for which $r_1$ and
$r_2$ quantum correlations vanish at $T=0$
}
\label{fig:zpd}
\end{center}
\end{figure}

Figure~\ref{fig:zpd} shows the domain for $r_1$ and $r_2$, i.e., phase diagram in the
plane $(r_1,r_2)$.
%
%
The domain consists of two regions,
$\Omega_0$ (which corresponds to the branches $Q_0$, ${\cal U}_0$, and ${\cal F}_0$)
and
$\Omega_1$ (which corresponds to the branches $Q_1$, ${\cal U}_1$, and ${\cal F}_1$),
separated by the boundary (\ref{eq:bound})
(solid line $r_1+r_2=2$ in Fig.~\ref{fig:zpd}).

\begin{figure}[t]
\begin{center}
\epsfig{file=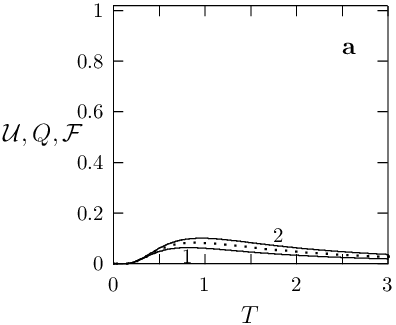,width=3.5cm}
\hspace{1mm}
\epsfig{file=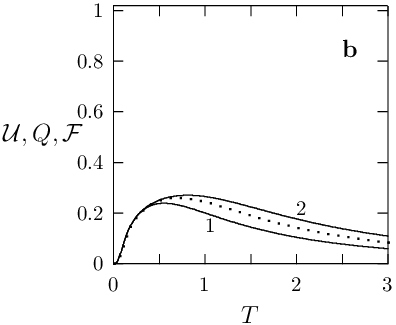,width=3.5cm}
\hspace{1mm}
\epsfig{file=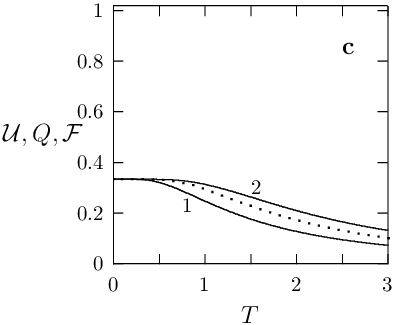,width=3.5cm}
\hspace{1mm}
\epsfig{file=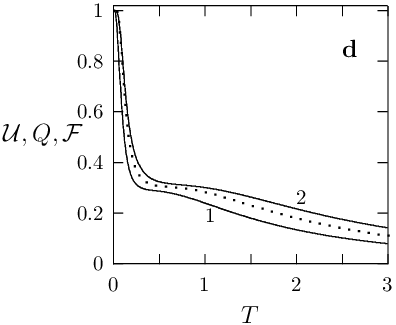,width=3.5cm}
\hspace{1mm}
\epsfig{file=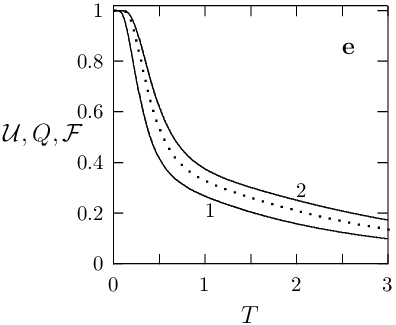,width=3.5cm}
\hspace{1mm}
\epsfig{file=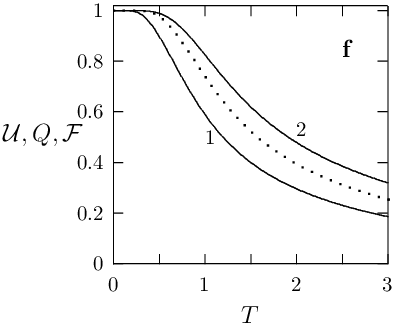,width=3.5cm}
\end{center}
\begin{center}
\caption{
$\cal U$ (solid line 1), $Q$ (dotted line), and $\cal F$  (solid line 2) vs $T$ for
$J_z=1$, $r_2=0$, and $r_1=1~(\bf a)$, $1.8~(\bf b)$, $2~(\bf c)$, $2.3~(\bf d)$,
$3~(\bf e)$, $5~(\bf f)$
}
\label{fig:r1r2_0}
\end{center}
\end{figure}

Consider the behavior of quantum correlations along the path $r_2=0$, i.e., on
abscissa axis.
At the origin of the Cartesian coordinates ($r_1=r_2=0$), the off-diagonal elements of
the density matrix $\varrho$, Eq.~(\ref{eq:varrho}), equal zero, the system is
classical and, therefore, quantum correlations are completely absent.
If $r_1$ starts to increase, quantum correlations appear as depicted in
Fig.~\ref{fig:r1r2_0}a, and they grow with increasing $r_1$, see
Fig.~\ref{fig:r1r2_0}b.
%
%
These curves have hill-like form which preserves up to $r_1=2_{-0}$.
We will refer to this behavior as behavior of type III.
It is characterized by zero quantum correlations at $T=0$ and nonzero at $T>0$.

To understand this somewhat unexpected behavior, consider off-diagonal elements
of the density matrix $\varrho$, Eq.~(\ref{eq:varrho}).
One off-diagonal element is $|v|=0$, and the other, for $T\to0$, has the form
\begin{equation}
   \label{eq:uT0}
   |u|\approx\frac{1}{2}\frac{1}{1+2\exp[(2-r_1)/T]}.
\end{equation}
It is clear that this quantity equals zero for $r_1<2$ in the low-temperature limit.
In other words, at zero temperature the density matrix becomes diagonal and therefore
all quantum correlations disappear.
Interestingly enough, the system loses quantumness at zero temperature, whereas the
same system contains nonclassical correlations for nonzero temperatures.

Observed behavior can also be established directly from the formulas for the quantum
correlations.
Indeed, for example, LQU on the abscissa in the $\Omega_0$ region is given, according
to Eqs.~(\ref{eq:Wxyz}) and (\ref{eq:U0U1}), as
\begin{equation}
   \label{eq:U0}
   {\cal U}_0(T)=\frac{2\sinh^2(r_1/2T)}{\cosh(r_1/T)+\exp(2/T)}.
\end{equation}
When the temperature goes to zero, LQU behaves as 
\begin{equation}
   \label{eq:U0a}
   {\cal U}_0(T)\approx\frac{1}{1+2\exp[(2-r_1)/T]}.
\end{equation}
Hence it follows that ${\cal U}_0(0)\equiv0$ for $r_1\in[0,2)$.
Thus, the quantum correlation on the segment shown in Fig.~\ref{fig:zpd} with a dotted
horizontal line on the abscissa, is completely suppressed at absolute zero
temperature.
This is valid for other two correlations $Q$ and $\cal F$, what is clear seen in
Fig.~\ref{fig:r1r2_0}~(a) and (b).

The third type of behavior of quantum correlations is radically different from the
cases shown in Fig.~\ref{fig:zUQFa}, where quantum correlations at $T=0$,
on the contrary, reach the maximum possible value equal to unity
(complete correlation).
Note that similar hill-like behavior of quantum discord was earlier observed, e.g., in
the spin systems with dipole-dipole interactions \cite{KY13}.

When $r_1$ reaches the value 2, a new qualitative change occurs in
behavior of quantum correlations, namely, they are equal to one third ($1/3$) at
zero absolute temperature.
This follows from Eq.~(\ref{eq:U0a}) and is clear seen in Fig.~\ref{fig:r1r2_0}c.
The value of quantum correlations here is not equal to zero or one at $T=0$,
correlations take an intermediate value.
This is the IV type of behavior of quantum correlations.

With a further increase in the value of $r_1$ while maintaining $r_2=0$, LQU goes to
another branch and becomes equal to
\begin{equation}
   \label{eq:U1}
   {\cal U}_1(T)=\frac{\cosh(r_1/T)+e^{2/T}-2e^{1/T}\cosh(r_1/2T)}{\cosh(r_1/T)+e^{2/T}}.
\end{equation}
At $r_1=2$, this equation also gives ${\cal U}_1=1/3$ in the low-temperature limit.
However, when $r_1>2$, the values of quantum correlations jump from 1/3 to one at zero
temperature and have monotonically decreasing shapes for $T>0$, as shown in
Fig.~\ref{fig:r1r2_0}d-f.
 
\begin{figure}[t]
\begin{center}
\epsfig{file=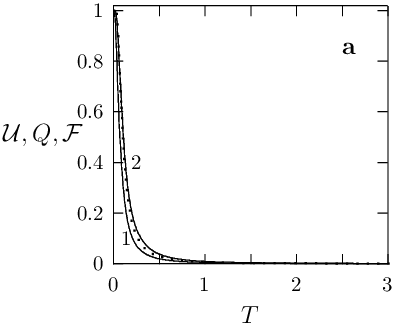,width=3.5cm}
\hspace{1mm}
\epsfig{file=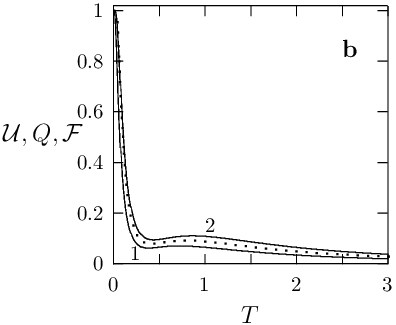,width=3.5cm}
\hspace{1mm}
\epsfig{file=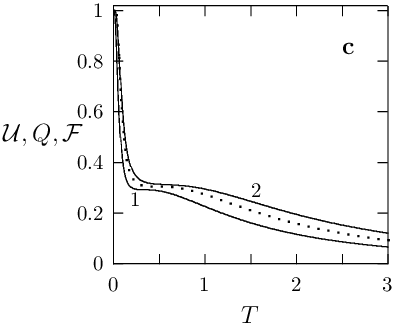,width=3.5cm}
\hspace{1mm}
\epsfig{file=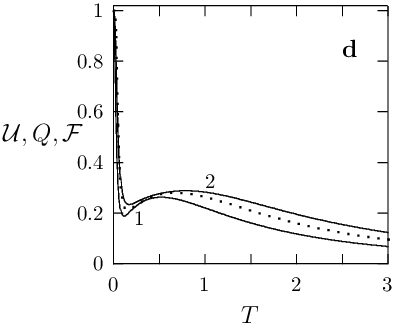,width=3.5cm}
\hspace{1mm}
\epsfig{file=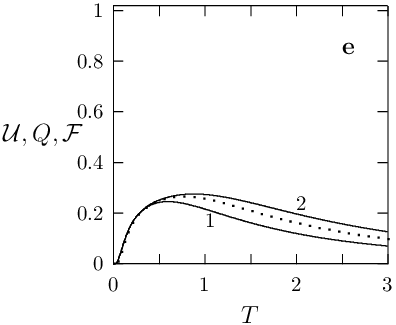,width=3.5cm}
\hspace{1mm}
\epsfig{file=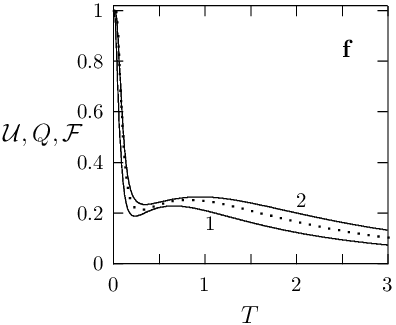,width=3.5cm}
\end{center}
\begin{center}
\caption{
$\cal U$ (solid line 1), $Q$ (dotted
line), and $\cal F$  (solid line 2) vs
$T$ for $J_z=1$, $r_2=0.1$, and $r_1=0~(\bf a)$, $1~(\bf b)$, $1.9~(\bf c)$,
$2~(\bf d)$, $2.1~(\bf e)$, $2.3~(\bf f)$
}
\label{fig:r1r2_0.1}
\end{center}
\end{figure}

Let us now turn to the evolution of quantum correlations for $r_2>0$.
Take, for example $r_2=0.1$, and let effective interaction $r_1$ increases from zero.
The transformations of the curve shapes are shown in Fig.~\ref{fig:r1r2_0.1}.
%
The first thing we observe is a qualitative change in behavior at $r1=0$, see
Fig.~\ref{fig:r1r2_0.1}a.
When $r_2>0$, the quantum correlations at $T=0$ are now equal to one rather than zero.
Otherwise, the curves repeat the behavior of the second and first types.

However there is an unexpected exception at $r_1=2.1$, where the maximum at $T=0$
suddenly disappears completely.
This is shown in Fig.~\ref{fig:r1r2_0.1}e.
To establish the reason for this behavior, consider again the structure of quantum
state $\varrho$ in the limit $T\to0$ under relation $r_1-r_2=2$.
For this purpose, take expressions for the diagonal, $a$ and $b$, and off-diagonal
matrix elements $|u|$ and $|v|$, Eqs.~(\ref{eq:avT}) and (\ref{eq:uv}).
Performing the necessary calculations, we obtain that the quantum state at zero
temperature is written as
\begin{eqnarray}
   \label{eq:varrho0a}
   \varrho_1=\frac{1}{4}
	 \left(
      \begin{array}{cccc}
      1&.&.&1\\
      .&1&1&.\\
      .&1&1&.\\
      1&.&.&1
      \end{array}
   \right).
\end{eqnarray}
Like (\ref{eq:varrho0}), the given state is classical and, hence, all quantum
correlations vanish at $T=0$ on the line $r_2=r_1-2$ (it is shown by dotted inclined
line in Fig.~\ref{fig:zpd}).
This phenomenon could be called the sudden death of quantum correlation at zero
temperature.

\begin{figure}[t]
\begin{center}
\epsfig{file=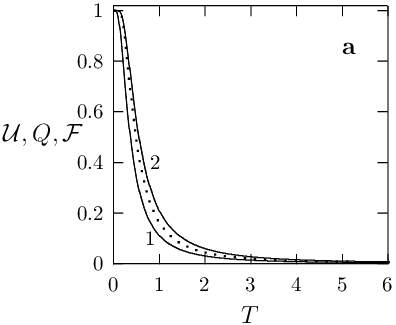,width=3.5cm}
\hspace{1mm}
\epsfig{file=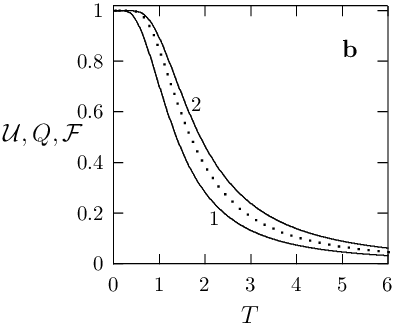,width=3.5cm}
\hspace{1mm}
\epsfig{file=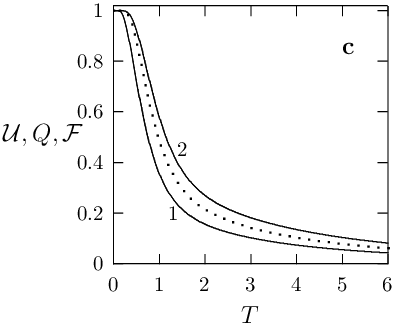,width=3.5cm}
\end{center}
\begin{center}
\caption{
Quantum correlations $\cal U$ (solid line 1), $Q$ (dotted
line), and $\cal F$  (solid line 2) versus temperature $T$ for $J_z=1$ and
$r_1=r_2=0.5~(\bf a)$; $r_1=0, r_2=2~(\bf b)$; $r_1=5, r_2=1~(\bf c)$
}
\label{fig:zz}
\end{center}
\end{figure}

In general, the following conclusion can be drawn.
Dependencies of quantum correlations near neighborhoods of the dotted polyline
(see Fig.~\ref{fig:zpd}) belong to the type II.
Away from this line, the quantum correlation curves decrease monotonically without
increase in any intermediate temperature range (type I of behavior).
As an illustration, we depicted the dependencies of LQU, discord, and LQFI in
Fig.~\ref{fig:zz} at a few randomly chosen point in the regions $\Omega_0$ and
$\Omega_1$ and on the boundary between them (see again Fig.~\ref{fig:zpd}).
%
%
Their behavior corresponds to type I.

\subsection{
Sudden change phenomena of quantum correlations
}
\label{subsect:QCSC}

\subsubsection{
$T>0$
}
\label{subsubsect:QCSCT}
According to Eqs.~(\ref{eq:Q0Q1}), (\ref{eq:min_U}), and (\ref{eq:min_F}), the
quantities $Q$, $\cal U$, and $\cal F$ are determined by choice from two alternatives.
This paves the way for the transitions of quantum correlations from one branch to
another during the evolution of the system in some parameters.
In catastrophe theory \cite{A92}, such abrupt qualitative transitions with a smooth
change in the control parameters are called sudden changes.

In the case under study, the situation is as follows. 
Since the boundary between the regions $\Omega_0$ and $\Omega_1$ does not depend on
temperature, transitions from one branch to another do not occur at
temperature changes.
The interaction parameters need to be changed.

\begin{figure}[t]
\begin{center}
\epsfig{file=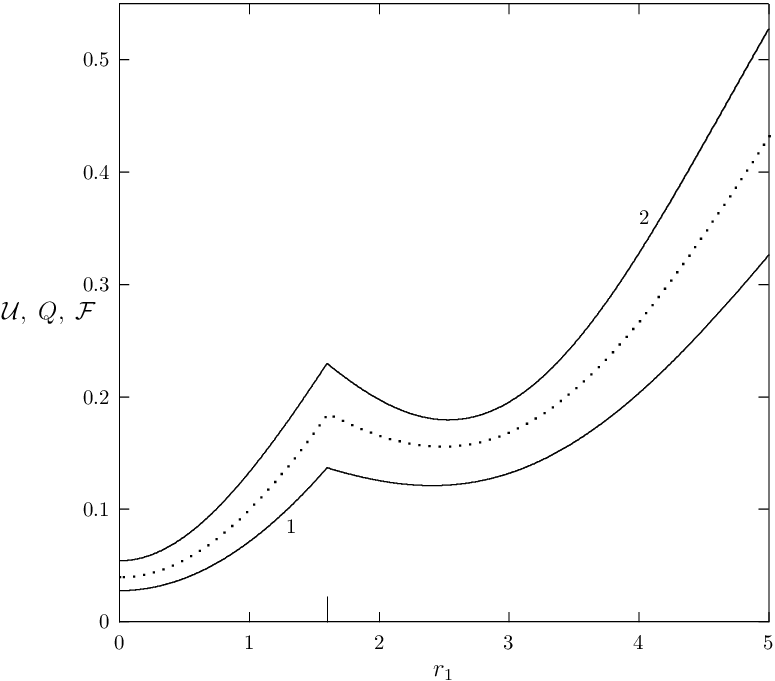,width=5.7cm}
\hspace{2mm}
\epsfig{file=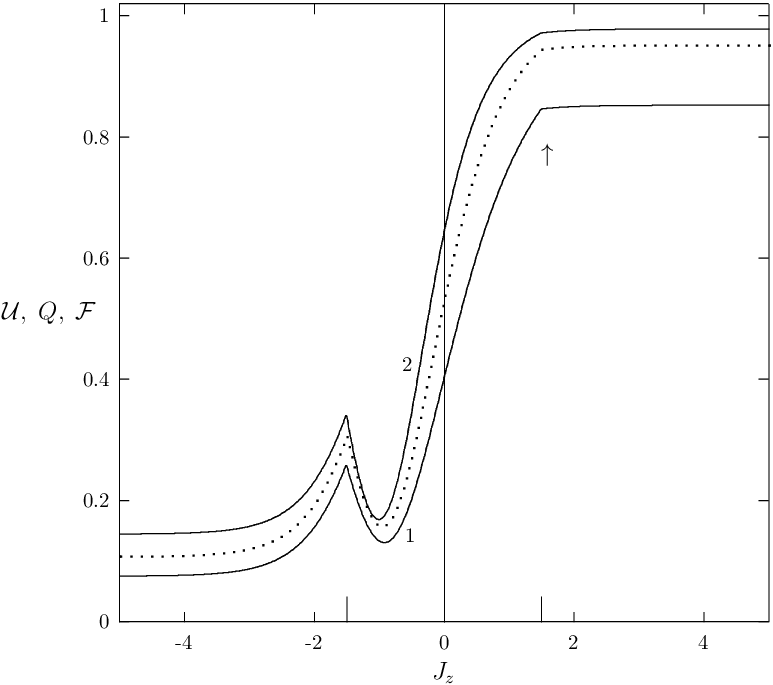,width=5.7cm}
\end{center}
\begin{center}
\caption{
$\cal U$ (solid line 1),
$Q$ (dotted
line), and $\cal F$  (solid line 2)
versus $r_1$ by $T=1.5$, $J_z=1$,
$r_2=0.4$  ($\bf a$) and
versus $J_z$ by $T=1$, $r_1=0.4$,
$r_2=2.6$ ($\bf b$).
Vertical arrow on the right panel indicates the position of fractures
(sharp bends) on curves for $J_z=1.5$
}
\label{fig:zUQF_J}
\end{center}
\end{figure}

Figure~\ref{fig:zUQF_J}a shows  the behavior of quantum correlations versus the
effective interaction parameter $r_1$.
%
%
It is clearly seen that all three dependencies have sharp maxima at $r_1=1.6$ (shown
by a longer bar on the abscissa), where the first derivatives of quantum correlation
functions with respect to $r_1$ undergo discontinuities of the first kind.
The point of sudden changes, $r_1=1.6$, lies at the boundary $r_1+r_2=2$ which
separates the regions $\Omega_0$ and $\Omega_1$ (see Fig.~\ref{fig:zpd}).

Two sudden changes can be seen in Fig.~\ref{fig:zUQF_J}b, where
quantum correlation dependencies are presented as functions of the longitudinal
interaction $J_z$.
Transitions occur when $J_z=\pm1.5$ (shown by two longer bars on the abscissa), which
follow from the condition $r_1+r_2=2|J_z|$.
One group of sudden changes, at $J_z=-1.5$, looks as cusp-like peaks.
Other sudden changes that occur when $J_z=1.5$ are much less pronounced.
They visible as weak fractures (kinks or bends), their position in the figure is
marked arrow pointing up.
All quantum correlation functions are continuous, but their first derivatives with
respect to the interaction $J_z$ undergo finite jumps (discontinuities).

In practice, experimental measurements of fractures and jumps can be used to estimate
the interaction parameters in the system.

\subsubsection{
$T=0$
}
\label{subsubsect:QCSC0}
The above picture take place for nonzero temperatures.
At $T=0$, the measures of quantum correlation coincide, ${\cal U}=Q={\cal F}$, and
they can undergo discontinuities themselves.
For example, when a completely cooled system evolves in $r_2$ along the trajectory
$r_1=0$ (see Fig.~\ref{fig:zpd}), LQU changes as
\begin{equation}
   \label{eq:QCr2}
   {\cal U}(r_1=0, r_2)=\cases{0,\quad  {\rm if}\ r_2=0 \cr 
	  1,\quad  {\rm if}\ r_2>0}.
\end{equation}
This can be seen by comparing Fig.~\ref{fig:r1r2_0}a and Fig.~\ref{fig:r1r2_0.1}a.
The same is valid for other two correlations, $Q$ and $\cal F$.

On the path $r_2=0$, LQU versus $r_1$ changes as
\begin{equation}
   \label{eq:QCr1}
   {\cal U}(r_1, r_2=0)=\cases{
	  0,\qquad\quad  {\rm if}\ r_1<2 \cr 
	  1/3,\qquad  {\rm if}\ r_1=2 \cr 
	  1,\qquad\quad  {\rm if}\ r_1>2}.
\end{equation}
Same for $Q$ and $\cal F$ as shown in Fig~\ref{fig:r1r2_0}b-d at $T=0$.

Such abrupt changes in quantum correlations can be attributed to quantum phase
transitions.

\section{
Conclusions
}
\label{sect:Concl}
In this paper, the two-qubit Heisenberg XYZ model with both antisymmetric
Dzyaloshinsky--Moriya and symmetric Kaplan--Shekhtman--Entin-Wohlman--Aharony
interactions has been considered at thermal equilibrium.
For it, we have examined the behavior of three measures of quantum correlation,
namely, the entropic quantum discord, local quantum uncertainty, and local quantum
Fisher information.
To classify the behavior of correlations, four qualitatively different types of curves
have been suggested.

Despite the different underlying concepts behind quantum correlations, the comparative
analysis showed good agreement between all measures.
This is clearly evidenced by all the graphic material presented in
Figs.~\ref{fig:zUQFa}, \ref{fig:z010-3}, and \ref{fig:r1r2_0}-\ref{fig:zUQF_J}.
That is, these measures are reduced to some one effective average measure.
The entropic quantum discord $Q$ could be taken as such an average measure, because it
lies between two other measures: ${\cal U}\le Q\le{\cal F}$.

Park \cite{P19} has found that for the ferromagnetic case ($J_z<1$;
cf. Figs.~\ref{fig:zUQFa} and \ref{fig:zpd}, right), the thermal discord in the
small $T$ region exhibits a local minimum due to the DM interaction.
In addition to this observation, we have established that local minima and maxima
can also appear in the antiferromagnetic case, and they are caused by the KSEA
interactions.
  
Next, all three measures as function of temperature are continuous and smooth.
On the other hand, at nonzero temperatures, quantum correlations can suddenly change
with a smooth change in the coupling parameters.
Such abrupt changes are accompanied by fractures in the curves of quantum
correlations.
Moreover, we have found that the quantum correlations themselves can exhibit
discontinuities at zero temperature.

Summing up, we conclude the following.
In spin systems with DM and KSEA interactions, very similar behavior is observed
for three different measures of nonclassical correlations:
for the entropic quantum discord and measures based on the Fischer and Wigner-Yanase
information.




\end{document}